\documentclass[journal, draftclsnofoot, one column, 12pt]{IEEEtran}

\usepackage{amsmath}
\usepackage{graphicx, subfigure}
\usepackage{amssymb}
\usepackage{cite}
\usepackage{url}
\usepackage[ruled,vlined,lined,ruled,linesnumbered]{algorithm2e}
\usepackage{relsize}
\usepackage[nodisplayskipstretch]{setspace}
\usepackage{longtable}
\usepackage{booktabs}
\usepackage{multirow}
\usepackage[subnum]{cases}
\usepackage[usenames, dvipsnames]{color}

\begin{document}
\title{Online Charging Scheduling Algorithms of Electric Vehicles in Smart Grid: An Overview}

\author{
Wanrong Tang, Suzhi Bi, and Ying Jun (Angela) Zhang

\thanks{This work is supported in part by General Research Funding (Project number 14200315) from the Research Grants Council of Hong Kong and Theme-Based Research Scheme (Project number T23-407/13-N). The work of S. Bi is supported in part by the National Natural Science Foundation of China (Project number 61501303) and the Foundation of Shenzhen City (Project number JCYJ20160307153818306).}

\thanks{W.~Tang is with the Department of Information Engineering, The Chinese University of Hong Kong.}

\thanks{S. Bi is with the College of Information Engineering, Shenzhen University. He is the corresponding author of this article. }

\thanks{Y. J. Zhang is with the Department of Information Engineering, The Chinese University of Hong Kong. She is also with Shenzhen Research Institute, The Chinese University of Hong Kong.}

}

\maketitle

\vspace{-1.8cm}

\section*{Abstract}
As an environment-friendly substitute for conventional fuel-powered vehicles, electric vehicles (EVs) and their components have been widely developed and deployed worldwide.
The large-scale integration of EVs into power grid brings both challenges and opportunities to the system performance.
On one hand, the load demand from EV charging imposes large impact on the stability and efficiency of power grid. On the other hand, EVs could potentially act as mobile energy storage systems to improve the power network performance,
such as load flattening, fast frequency control, and facilitating renewable energy integration.
Evidently, uncontrolled EV charging could lead to inefficient power network operation or even security issues. This spurs enormous research interests in designing charging coordination mechanisms.
A key design challenge here lies in the lack of complete knowledge of events that occur in the future. Indeed, the amount of knowledge of future events significantly impacts the design of efficient charging control algorithms.
This article focuses on introducing online EV charging scheduling techniques that deal with different degrees of uncertainty and randomness of future knowledge.
Besides, we highlight the promising future research directions for EV charging control.

\section{Introduction}

\begin{figure*}
\centering
\includegraphics[width=1\textwidth]{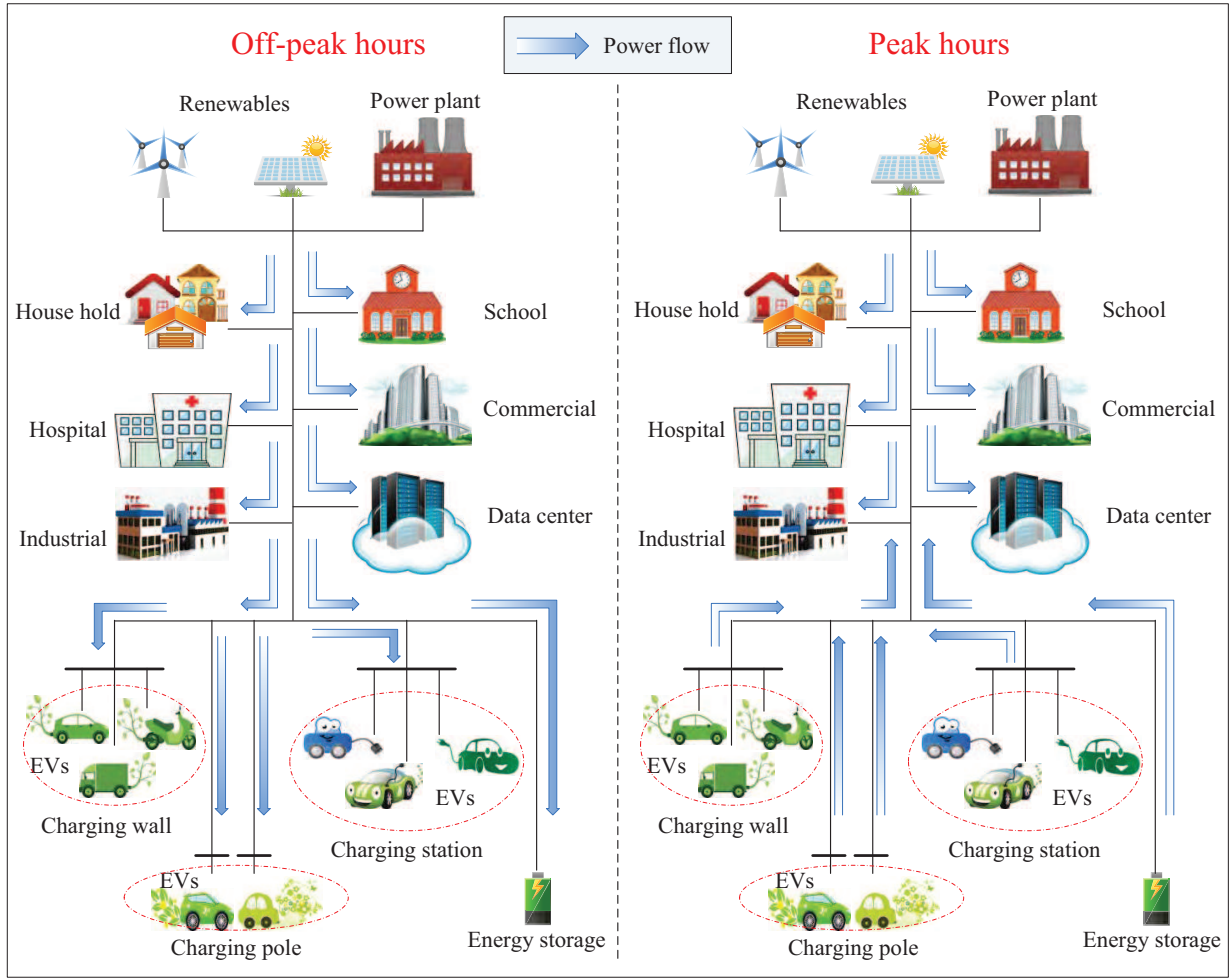}
\caption{An illustration of the applications of EVs at the time of peak hours and off-peak hours of base load consumptions.}
\label{fig:introduction1}
\end{figure*}

Electric vehicles (EVs) are referred as the vehicles that are powered fully or partially by electricity energy.
In general, the rechargeable battery of an EV can be charged from an external source of electricity through wall sockets, and also discharged to an external energy storage or power grid.
Compared with conventional fuel-powered vehicles, EVs produce very little air pollution upon their use.
In addition, the environmental benefits of EVs are magnified when they are powered by new and clean renewable energy sources, such as solar and wind power.
As such, a wide range of countries have pledged billions of dollars to fund the development of EVs and their components in an attempt to replace the conventional vehicles.
According to the recent analysis from the Centre for Solar Energy and Hydrogen Research, the demand of EV accounts for a total global market of more than 740,000 EVs in early 2015 \footnote{ From ZSW Centre for Solar Energy and Hydrogen Research, Mar. 2015, available at http://www.zsw-bw.de/en/support/news/news-detail/mehr-als-740000-autos-weltweit-fahren-mit-strom.html.}. In the next 50 years, the number of vehicles in operation is expected to increase from 700 million to 2.5 billion, where EVs will constitute a major part of them.

The fast increasing adoption of EVs brings both challenges and opportunities to the power grid.
On one hand, the massive load caused by the integration of EVs into the power grid raises concerns about the potential impacts to the operating cost, voltage stability and the frequency excursion at both generation and transmission sides.
On the other hand, EVs can be used as a new type of mobile energy storage systems that can serve many purposes.
With adequate energy stored in the batteries of EVs, the bidirectional charging and discharging control has extensive applications in the microgrids/distribution networks, such as load flattening, peak shaving, frequency fluctuation mitigation and improving the integration of renewable sources.
For instance, Fig.~\ref{fig:introduction1} illustrates the use of EVs for load flattening in a power gird.
During the off-peak hours, EVs can act as loads to withdraw and store electricity from the main grid.
During the peak hours, however, the EVs can release the stored energy back to the grid to meet the high demand of other electricity consumers. Overall, the use of EVs flattens the power profile over time and improves the stability of the entire power system.

In both cases, uncontrolled EV charging/discharging will lead to inefficient system operation or even severe security problems.
To mitigate the negative effects and enjoy the benefit of EV integration, it is critical to develop effective charging/discharging scheduling algorithms for efficient grid operation.
In practice,
a key design challenge of charging scheduling algorithms lies in the randomness and uncertainty of future events, including the charging profiles of EVs arriving in the future, future load demand in the grid, future renewable energy generation, etc.
Therefore, it is necessary to develop online charging/discharging algorithms to cope with different degrees of uncertainty when making real-time decisions.
Besides, the large-scale charging of EVs requires low-complexity control mechanisms to reduce the operating delay and the capital cost of equipment investment.
In this article, we introduce various online EV charging control mechanisms to enhance the efficiency and stability of power networks.
We discuss different online algorithms under different types of knowledge of future data, including the estimation of near-future random data, the mean, variance, and distribution, etc.
Specially, we explore some unique features of the charging behaviors of EVs to improve the general online algorithms for better performance and lower complexity.
We also notice that there are existing surveys on the energy management strategies of EVs proposed up to 2012 \cite{Panday2014}. In comparison, we not only update the state-of-the-art EV energy management technologies, but also focus on introducing the design of online charging scheduling algorithms.

This article is organized as follows. We first provide the basic model of online stochastic EV charging control. Then, we introduce the most up-to-date methods to tackle the EV charging scheduling problems under different degrees of knowledge of future information.
At last, we discuss the future research directions for online stochastic EV charging control in some interesting applications and conclude the article.

\section{Basic Model of Realtime Stochastic EV Charging Control}
\begin{figure*}
\centering
\includegraphics[width=0.5\textwidth]{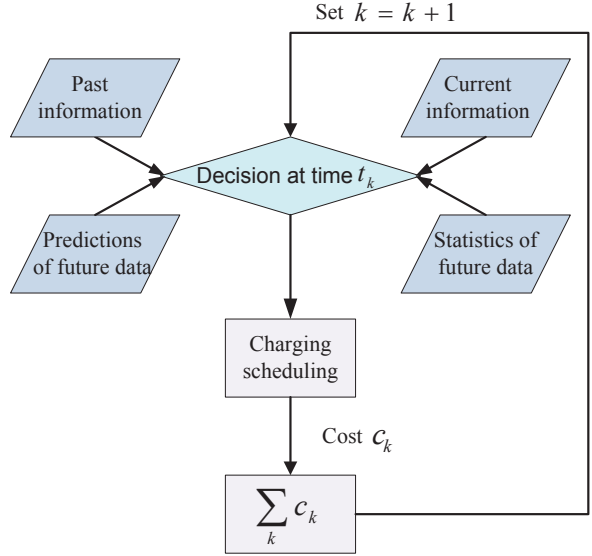}
\caption{The illustration of the online EV charging scheduling process.}
\label{fig:model1}
\end{figure*}

The online EV charging problem assumes that, at any time, the scheduler only knows the causal information, i.e., the information revealed so far. For instance, a charging facility, e.g., a charging station, only knows the charging profiles of the EVs that have arrived as well as the load demand and renewable energy generation in the grid up to the current time. Based on the causal information, the scheduler makes a charging decision, i.e., the current charging rates of all arrived EVs.
Notice that a past decision that has already been implemented cannot be reversed in the future.
In the following, we specify the elements of online EV charging control.

\emph{Event driven:}
In practice, an EV can arrive or depart at any time instant. As such, the charging schedule is a function of continuous time, which involves infinite number of control variables in the EV charging problem.
In fact, it has been shown in \cite{tang2014online} that the charging schedules only need to be updated at the time when an ``event'' occurs, such that the current system state changes. For instance, an event can be the arrival or departure of an EV, or the change of base load or electricity price.
Specifically, we denote by $t_1, t_2, \cdots, t_k$ the time when events $1, 2, \cdots, k$ occur, respectively. In general, the time length between time $t_k$ and time $t_{k+1}$ is a variable, which is decided by the random events.

\emph{System time:}
The system time horizon can be either finite or infinite. In practice, an EV charging schedule is optimized over a finite time horizon of from several hours to several days, while the length of a time slot is often in the order of minutes.
The system time horizon can be regarded as infinite when it is much longer than the length of a time slot, e.g., several years.

\emph{Causal information:}
In the realtime scenario, only the past and current information is known by the charging scheduler.
For instance, at any time slot, a charging station only knows the charging demands and departure deadlines of the EVs that arrive at or before current time, the past and current base load and renewable energy, etc.

\emph{Random data:}
Due to the assumption of causality of knowledge, the non-causal information about future events appears uncertain and random. The randomness mainly comes from the following aspects:
1) charging profiles of EVs that arrive in the future, including arrival, departure, charging demand, and individual charging constraints,
2) the future load demand in the grid by, for example, residential buildings, factories, schools, hospitals, commercial buildings, data centers,
etc. 3) future renewable energy generations from, for example, solar, wind, and hydro-electric plants,
4) future prices including electricity price and regulation service price.

\emph{Knowledge of future data:}
Based on the historical data, the scheduler may have some predictions on the future data, including the near-future predictions or the statistics such as the mean, variance and distributions.

\emph{Objective:} The objective EV charging control varies depending on the standpoint we choose to take.
From EV owners' viewpoint, the objectives could be charging demand satisfaction (i.e., fulfilling the EVs' charging demands before their specified deadlines), charging cost minimization, or profit maximization by selling power to the power grid.
On the other hand, the objective of a utility owner could be energy cost minimization, load flattening/shaping, peak shaving, frequency regulation, and voltage regulation.
In general, the objective of a charging scheduling problem can be expressed as a cost function to be minimized.

Based on the above definitions, the process of a general online EV charging scheduling can be described as Fig.~\ref{fig:model1}.
At time $t_k$, the scheduler makes a decision based on the causal information and the possible predictions/statistics of future random data, and then induces a cost, denoted by $c_k$. The process repeats until the system time ends. We denoted by $T$ the total number of times that the decisions are made.
The charging decision and random data revealed at time $t_k$ are denoted by $\textbf{x}_k$ and $\boldsymbol{\xi}_k$, respectively.
The charging decisions and random data revealed from time $t_1$ to $t_k$ are denoted by $\textbf{x}_{1:k}$ and $\boldsymbol{\xi}_{1:k}$ respectively.
Specially, the cost at time $t_k$ is a function of the charging decisions and random data revealed up to time $t_k$, i.e., $c_k = f(\textbf{x}_{1:k}, \boldsymbol{\xi}_{1:k})$.
Notice that the charging decisions depend on the the knowledge of the random data in the future.
In the next section, we will introduce the methodologies of online EV charging scheduling based on the knowledge of future random data and discuss their performance respectively.

\section{Stochastic Control Techniques of EV Charging}
\begin{figure*}
\centering
\includegraphics[width=0.9\textwidth]{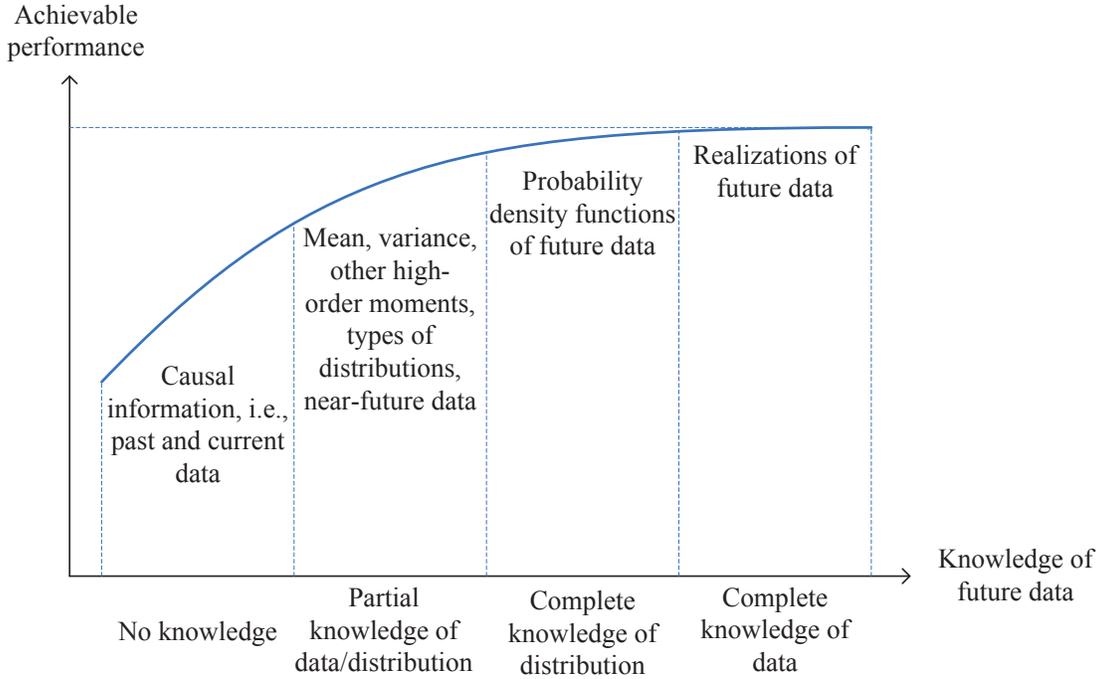}
\caption{The illustration of the spectrum of future knowledge.}
\label{fig:knowledge}
\end{figure*}

The knowledge of future random data is rather different in different applications.
Fig.~\ref{fig:knowledge} illustrates the spectrum of future knowledge.
As shown in Fig.~\ref{fig:knowledge}, the most ideal case is when the complete knowledge of the future data is known.
That is, the charging scheduler knows all the realizations of the future data before the beginning of system time.
Then, the stochastic scheduling problem for EV charging becomes a deterministic problem, which is much easier to tackle with deterministic algorithms.
Another extreme case is when absolutely no information about future data is known by the online charging scheduler. Then, the scheduler makes decisions based only on the data that has already revealed.
In between, the more general cases are that the scheduler has knowledge of some statistical information or short-term predictions of future data.
For instance, the statistical information of the EV traffic patterns could often be acquired through historic data, while the near-future data of renewable energy generation, e.g., the solar and wind power, can be predicted with high precision.

\subsection{Methodologies with Complete Knowledge of Future Data}
We first consider the case that the complete knowledge of data is known beforehand.
In this case, the random data at all times of making decision $\boldsymbol{\xi}_{1:T}$ become deterministic. Then, the stochastic EV charging problem is reduced to a deterministic problem, which is often referred to as \emph{offline problem}. The optimal solution to the offline problem is called \emph{optimal offline solution}, and the algorithm adopted to solve the offline problem is called \emph{offline algorithm}.
Specifically, the optimal solution, denoted by $\textbf{x}^*_{1:T}$, is calculated by
\begin{equation}
\textbf{x}^*_{1:T} = \arg \min_{\textbf{x}_{1:T}} \sum_{t=1}^T f(\textbf{x}_{1:t}, \boldsymbol{\xi}_{1:t}).
\end{equation}
Note that offline problem is deterministic and in general easier to handle than the online problem.
The optimal offline solution is not achievable in practice due to the unrealistic assumption of complete future information. Instead, it is often used as a benchmark to evaluate other online charging scheduling methods.

\subsection{Methodologies with Zero Knowledge of Future Data}
When no information about the future data is known, the charging scheduling algorithm makes decisions based on only the causal information available to the scheduler.
A key feature of the online algorithm is that the performance is generally evaluated in the worst case scenario, as no statistics of data could be leveraged to evaluate the average cost.
A standard metric to evaluate the worst-case performance of an online algorithm is \emph{competitive ratio}, defined as the maximum ratio between the cost achieved by an online algorithm and that achieved by the optimal offline algorithm over all possible input sequences (e.g., the EV arrival patterns, charging demands, and base load variations).
Let $\Phi$ be an online algorithm or policy, $\Pi$ be the set of all feasible policies, and $\textbf{x}_{1:t}^\Phi$ be the decision at time $t_1, \cdots, t_i$ under algorithm/policy $\Phi$.
Then, the optimal competitive ratio of policy $\Phi$ is calculated by
\begin{equation}
\min_{\Phi \in \Pi} \max_{\xi_{1:T}} \frac{\sum_{i=1}^T f(\textbf{x}_{1:i}^\Phi, \boldsymbol{\xi}_{1:i})}{\sum_{i=1}^T f(\textbf{x}_{1:i}^*, \boldsymbol{\xi}_{1:i})}.
\end{equation}
To minimize the competitive ratio, there are three main ideas to design competitive online algorithms for EV charging problem.
\begin{itemize}
  \item \emph{Classic online scheduling algorithms:} There exist many classic online scheduling algorithms that were proposed to solve problems other than EV scheduling, such as computing job scheduling and industrial process optimization. Some well-known methods include earliest deadline first (EDF) algorithms, the least laxity first (LLF) algorithm and optimal available (OA) algorithm \cite{he2012optimal}.
      When applied to EV charging, the EDF always charges the EV with earliest departure time first, the LLF schedules the EV with least laxity (i.e., the parking time length minus the shortest time length of fulfilling charging), and the OA solves the problem by assuming that no random data (or EVs, base load, etc) will be released in the future. In practice, however, the direct extension of these algorithms to EV charging may yield poor performance due to the special features of EV charging problem, e.g., the bursty and time-varying nature of EV arrivals. These classic algorithms often need modifications to fit in the structure of EV charging problems. Sometimes, the algorithms are combined with pricing and other control schemes, e.g, admission control \cite{chen2012iems}.
  \item \emph{Solution-structure based algorithms:} These algorithms are designed by exploring the structures of the optimal offline solution, given that it is easy to obtain.
      Indeed, exploring the offline solution structure is often used as the first step of online algorithm design. By observing the optimal offline solution, we try to fathom its solution structure. For example, when the objective function in the offline problem is an increasing convex function of the total load from EV charging and other elastic load, an optimal solution to the offline problem always tends to flatten the total load profile over time as much as possible \cite{tang2014online}\cite{gan2012optimal}\cite{tang2016onlineMPC}. This leads to the design of online algorithms that charge the EVs neither too fast nor too slowly to reduce the fluctuation of the total load.
  \item \emph{Data-mining/data-driven based algorithms:}
  The data-mining/data-driven based algorithms are designed by mining the revealed data and analyzing the statistics.  The statistics of the available data include the cross-correlation, auto-correlation and partial auto-correlation, etc. Typical data-mining/data-driven based algorithms include genetic algorithms, neural networks and fuzzy rule-based systems. In general, the data-mining/data-driven algorithms are more suitable for the case where the structure of system model can not be easily determined using empirical or analytical approaches \cite{xydas2016a}.
\end{itemize}

An efficient design of online EV charging scheduling is often a combination of the above methods.
For instance, assuming that the cost function is quadratic with the load, we get the insight that the  optimal offline solution should exhibit a load-flattening structure.
Meanwhile, we notice that the classic online algorithm OA only flattens the load demand revealed at current time but underestimates the load demand revealed in future.
In practice, the pattern of random EV arrivals often has some peaks.
By taking into account the possible peak arrivals of EVs in the future, an online algorithm named ORCHARD that speeds up the charging rate of OA by a proper factor is proposed in \cite{tang2014online}, which effectively reduces the possible peak load in the future.
As a result, the competitive ratio of online algorithm ORCHARD is shown to be 2.39, which is significantly better than that achieved by the original OA algorithm, i.e., 4.


Notice that most existing online algorithms for EV charging scheduling problem are deterministic, i.e., fixed decision output as a function of causal information input. A promising method to improve the worst-case performance of existing deterministic online algorithms is to apply randomized online algorithm.
A randomized online algorithm is a random strategy over a set of deterministic online algorithms based on a probability distribution.
For instance, the key idea of the algorithm designed in \cite{tang2014online} is to speed up the processing rate (charging rate) of OA by a factor, where the factor is a fixed constant.
A possible randomized online algorithm is to set the factor as the random variable which follows a certain probability distribution.
In general, randomized online algorithms have better worst-case performance than the deterministic online algorithms. However, the difficulty often lies in the setting of the probability distribution of a random algorithm.

\subsection{Methodologies with Partial Knowledge of Future Data}
In practice, some partial knowledge of future data, e.g., from the prediction of future data, is available in the design of online algorithms. For instance, power generation and load prediction algorithms are now important components of most modern smart grid. Indeed, the wind speed can be well-predicted by combining probability and fuzzy systems concepts \cite{zhang12design}.
For the EV charging problem, EV charging profiles can be predicted based on the past data collected and reservations made by the EV users in advance.
In general, statistical-modeling based algorithms are often applied for data prediction, e.g., artificial neural network (ANN), EV user classification, and other Machine Learning (ML)-based methods \cite{majidpour15fast}.
By incorporating the near future estimation, online algorithms could be designed to neglect some unrealistic worst cases and improve performance based on the partially-known future.

\subsection{Methodologies with Knowledge of Statistical Information}
In this section, we discuss the case where the future data is not known, but its statistical information can be estimated based on the historic data.
The estimation of the future random processes mainly includes the estimation of the moments (e.g., mean as the first-order moment and variance as the second-order moment) and the estimation of probability distributions (i.e., moments of all orders).
When the scheduler has the knowledge of probability distributions of random data, i.e., probability density functions (PDF), algorithms based on dynamic programming can be applied.
When the number of times of making decision is finite, the problem can be solved by backward induction method or Monte Carlo sampling techniques \cite{leou2014stochastic}.
When the number of times of making decision goes to infinity, the problem can be formulated as an infinite-time horizon dynamic programming or a Markov Decision Process (MDP).
Specifically, we denote by $s_k$ the \emph{system state} at time $t_k$, e.g., the current charging demand of individual EV, the base load, and electricity price, etc.
The \emph{action} is the charging decision at time $t_k$, i.e., $\textbf{x}_k$.
Then, the online EV charging problem is that at time $t_k$, the decision maker chooses an action $\textbf{x}_k$ that is available in current state $s_k$. The process responds at the next time step by randomly moving into a new state $s_{k+1}$ following a known distribution, 
and then returns a corresponding cost-to-go, denoted by  $v_k(s_k)$.
Specifically, the optimal cost-to-go, denote by $v_k^*(s_k)$ at time $t_k$, satisfies the following Bellman's equation \cite{zhang2016profit}
\begin{equation}\label{eq:bellman}
v_k^*(s_k) = \min_{\textbf{x}_k} f(\textbf{x}_{1:k}, \boldsymbol{\xi}_{1:k}) + \alpha \sum_{s_{k+1}} P(s_k, s_{k+1}) v_{k+1}^*( s_{k+1} ),
\end{equation}
where $\alpha$ is a discount factor and $P(s_k, s_{k+1})$ is the transition probability from $s_k$ to $s_{k+1}$.
Note that the EV charging process is featured by the battery memory.
When formulating the EV charging problem as a Markov Decision Process (MDP), the system state could be defined as the energy levels of the battery stored in the EV or the renewable power supplied in the system. The transition probability could be estimated by the historic data of the renewable power and EV charging demands.
There are several standard algorithms to solve the MDP problem, e.g., value iteration, policy iteration, modified policy iteration, and prioritized sweeping, etc.
When the statistic information of the random data is not clear, Q-learning algorithm could be adopted to solve the MDP problem.
Note that the EV charging problem often contains a continuous space of system state, e.g., the energy level of battery and the electricity price, and a continuous space of action, i.e., the charging rate.
The existing research often uses discrete Bellman's equation to  model the EV charging problem \cite{donadee2014stochastic}\cite{zhang2016profit}, which can lead to prohibitive computation complexity.
On the other hand, as the fast integration of EVs into the power grid, the large scale of EVs could also bring the issue about the curse of dimensionality.
To reduce the computational complexity, approximate (stochastic) dynamic programming (ADP) methods could be adopted  \cite{donadee2014stochastic}.

In most cases, it is hard to accurately estimate the complete probability density function of the random data based on the historic data.
A more practical prediction of data statistics is the low-order moment, e.g.,  the mean and the variance, as it requires much fewer data samples than to accurately characterize the full probability distribution.
Then, advanced techniques from robust optimization could be adopted to tackle the online problems with partial statistic information.
Since the first-order moment is the simplest to estimate compared with other statistics, a lot of works make use of the mean instead of high-order information.
Specifically, Model Predictive Control (MPC) method is one common approach to handle online problems with the knowledge of the expected values of random data.
To address a wide range of uncertainties and variability, MPC based charging scheduling algorithm replaces all future data, e.g., renewable energy, base load, arrival rate and charging load demand of EVs, by their expected values and thus reduce stochastic problem to a deterministic problem.
A well-accepted metric to valuate MPC based charging scheduling algorithm is Value of the Stochastic Solution (VSS), which evaluates the optimality gap between the optimal solution to \eqref{eq:bellman} by requiring the distributions of $\xi$ and the solution from MPC based algorithm by replacing $\xi$ with the means \cite{tang2016onlineMPC}.
In practice, the statistics of EV arrival process often exhibit periodicity. For example, the arrival rate of the residential EV charging demand could have a periodicity, where the period is one day
\footnote{X.~Zhang and S.~Grijalva, ``An Advanced Data Driven Model for Residential  Electric Vehicle Charging Demand,'' technique report, Georgia Institute of Technology, 2015.}. The daily travel patterns are also likely to exhibit periodicity based on the National Household Travel Survey (NHTS) 2009
\footnote{The National Household Travel Survey (NHTS) 2009 gathers information about daily travel patterns of different types of households in 2009, and shows that the daily travel statistics (e.g., Average Vehicle Trip Length, Average Time Spent Driving, Person Trips, Person Miles of Travel) are very similar for each weekday or weekend.}.
Accordingly, the periodicity of EV random arrival process can facilitate the prediction of EVs' arrivals to improve the performance.
For instance, \cite{tang2016onlineMPC} shows that the MPC based algorithm could be made more scalable if the random process describing the arrival of charging demands is first-order periodic.
Besides, another scenario is to assume that the random data comes from a population that follows a known probability distribution, where the typical parameters, i.e., mean, variance, etc, are unknown.
These parameters can be estimated by elementary statistical methods and made more accurate by sensitivity analysis.
For instance, the recent studies on the real-world data verify the hypothesize that the aggregate arrival rates
of EVs follow a Poisson distribution \cite{alizadeh2014a}.

For the ease of reference, we summarize the methodologies to design online EV charging scheduling algorithms in Table.~\ref{table:knowledge}.
For the case with complete knowledge of distribution, the algorithms are likely to induce high computational complexity. In this case, exploiting special solution structure may lead to a greatly reduced computational cost. For example, a threshold-based charging algorithm is developed in \cite{zhang2016profit}.
For the case with partial knowledge of statistics, it is of high interest to improve the performance of sub-optimal scheduling solution. One possible solution is to combine online/stochatic learning techniques and robust optimization to improve the performance of the algorithm.

\begin{table*}
\caption{Summary of knowledge of future information and common methodologies}
\label{table:knowledge}
\centering
\small
\begin{tabular}{|c|c|c|}
\hline
\multirow{2}{*}{\emph{Knowledge categories}} & \emph{Future information} & \multirow{2}{*}{\emph{Methodologies}}\\
& \emph{known by Scheduler} &  \\ \hline
\multirow{3}{*}{\textbf{complete knowledge of data}} & \multirow{3}{*}{realizations}  &   linear programming, convex optimizations, \\
 &   &  graph algorithms, greedy algorithms, \\
 &     &   approximation algorithms, heuristic algorithms \\ \hline
\multirow{2}{*}{\textbf{complete knowledge of distribution}} & probability density   &   dynamic programming,  Markov decision process,       \\
   &   functions  &   stochastic dynamic programming, Monte Carlo sampling  \\ \hline
\multirow{3}{*}{\textbf{partial knowledge of distribution}} &  first-order moments  &  model predictive control \\ \cline{2-3}
    &   high-order moments   &  robust optimizations \\ \cline{2-3}
    &    types of distributions &  parametric methods  \\ \hline
\multirow{2}{*}{\textbf{partial knowledge of data}} & \multirow{2}{*}{near-future data}  &   Markov models, time series, \\
                                      &       &   machine-learning based algorithms \\ \hline
\multirow{3}{*}{\textbf{no knowledge of data}} & \multirow{3}{*}{zero}  &   classic online scheduling algorithms, \\
 &   &  solution-structure algorithms, \\
   &       &   data-mining/data-driven based algorithms \\ \hline
\end{tabular}
\end{table*}

\section{Performance Evaluation}
\begin{figure*}
\centering
\includegraphics[width=0.7\textwidth]{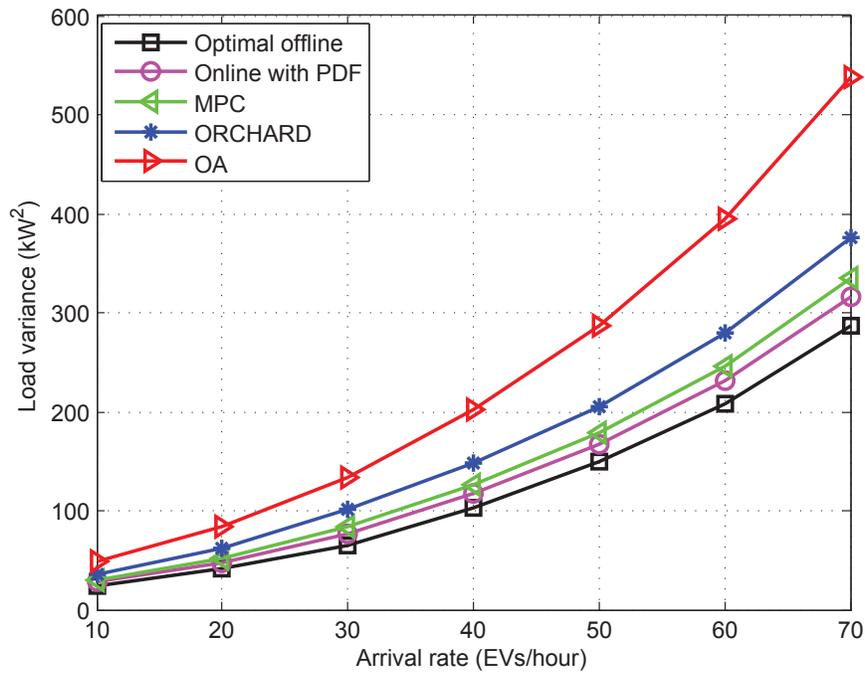}
\caption{Load variance of five algorithms over arrival rate at peak hours in scenario 1, where EVs act as load demands.}
\label{fig:var1}
\end{figure*}

\begin{figure*}
\centering
\includegraphics[width=0.7\textwidth]{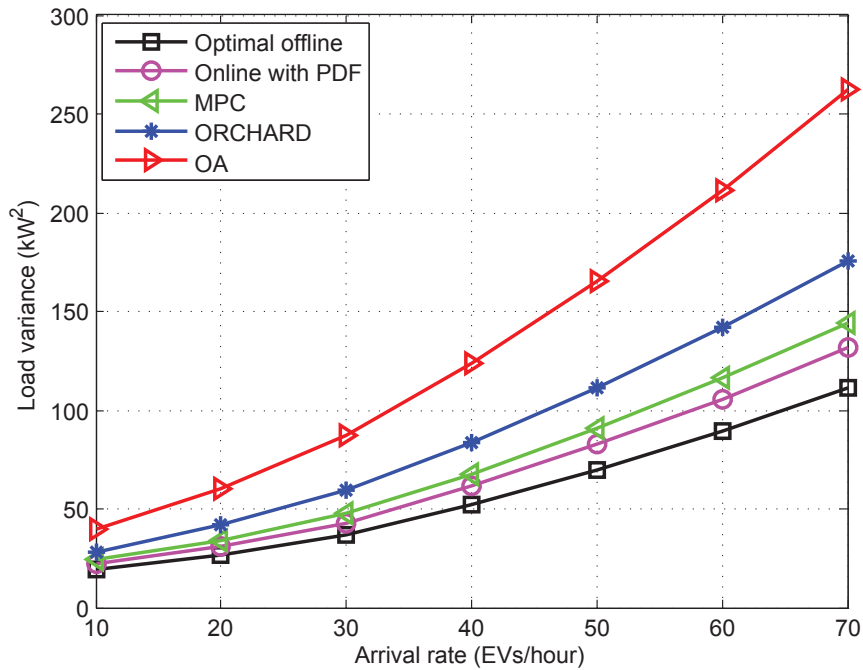}
\caption{Load variance of five algorithms over arrival rate at peak hours in scenario 2, where EVs act as both load demands and power sources.}
\label{fig:var2}
\end{figure*}

In this section, we evaluate the performance of the methodologies discussed above.
The system time is set to be $24$ hours, and the length between two adjacent times of making decision is set to be 10 minutes.
Suppose that the EV arrivals follow a Poisson distribution and the parking time of each EV follows an exponential distribution \cite{alizadeh2014a}.
Their charging demand follows an uniform distribution.
For the traffic patterns, we set two peak periods, i.e., $12:00$ to $14:00$ and $18:00$ to $20:00$,
which match with the realistic vehicle trips in National Household Travel Survey (NHTS) 2009.
We investigate two scenarios where the EVs serve for different purposes. In scenario 1, EVs act only as the consumers that require to satisfy the charging demand. 
In scenario 2, EVs act as not only consumers but also power suppliers, where EVs could be charged/discharged from/to the grid.
For both scenarios, the objective function is to minimize the variance of total load, which consists of the load from EV charging and the inelastic base load. The minimization of load variance in effect reduces system power losses and improves voltage regulation \cite{sortomme2011coordinated}.
Specifically, we choose the following algorithms listed in a decreasing order of the amount of future data knowledge.
\begin{enumerate}
  \item Optimal offline algorithm: the complete knowledge of the random data is assumed to be known. Specifically, we adopt interior point method in CVX\footnote{M.~Grant and S.~Boyd, CVX: Matlab Software for Disciplined Convex Programming [Online]. Available: http://cvxr.com/cvx Mar. 2013, Version 2.0 (beta).} to compute the optimal offline solution.
  \item Online algorithm with PDF: the complete knowledge of distributions of random data are assumed to be known. Specifically, we adopt sample average approximation (SAA) method as the online algorithm with PDF.
  \item Online algorithm MPC \cite{tang2016onlineMPC}: the expected values of the random data are assumed to be known.
  \item Online algorithm with no knowledge of future information: ORCHARD \cite{tang2014online} and  OA \cite{he2012optimal} : no future information is assumed to be known.
\end{enumerate}
For both scenarios, we plot the load variance of the five algorithms by increasing the arrival rates during the peak hours, as shown in Fig.~\ref{fig:var1} and Fig.~\ref{fig:var2}.
Both figures show that the optimal offline algorithm always produces the lowest load variance among the five algorithms.
Meanwhile, the online algorithm with PDF achieves lower cost than the MPC algorithm with prediction of means, and both algorithms follow closely to the optimal offline algorithm.
We also notice that online algorithm ORCHARD and OA produce higher load variance than the other three algorithms, since they assume no predictions nor non-causal information of the random data.
Between them, ORCHARD significantly outperforms OA, where the OA algorithm performs poorly especially under high peak arrival rate.
For all five algorithms, it can be easily observed that the load variance of scenario 2 depicted in Fig.~\ref{fig:var2} is much smaller than that of scenario 1 depicted in Fig.~\ref{fig:var1}, which demonstrates the effectiveness of using EVs as mobile energy storage to flatten the system load profiles.

\section{Future Research Directions}
The online algorithm design for EV charging scheduling contains rich research problems with different applications of EVs. In this section, we highlight several interesting research topics we deem particularly worth investigating.

\subsection{Economic Incentive Design}
The major challenge of the online charging algorithm design is the uncertainties from the behavior of EV users.
A promising solution is to introduce economic incentive schemes to encourage more users to arrive at the charging station during the off-peak hour of base load consumptions and less during the peak hours, so that the total load demand is flattened over time. Equivalently, pricing method can be used to adjust the EVs' charging demand over time.  For instance, distribution locational marginal pricing method could be adopted to alleviate congestion induced by EV loads \cite{li2014distribution}. Besides, the scheduler can also offer financial compensation to those users who are willing to make reservations day-ahead, park the EV for a longer time, or tolerate charging delay after the specified parking time. Through optimizing the pricing schemes, the scheduler maximizes its overall utility, e.g., its profit defined as the revenue minus the operating cost and the cost on offering the incentives. The joint design of pricing scheme and online EV scheduling is also a promising yet challenging topic to investigate, considering the complex correlations between the pricing and the EV user profiles, including arrival rates, parking time and charging demand.

\subsection{Online/stochastic Learning of Random Data }
As shown in Fig.~\ref{fig:var1} and Fig.~\ref{fig:var2}, the accurate knowledge of future data can lead to significant performance improvement of online algorithms. Currently, most studies on online scheduling design assume perfect knowledge of (partial) future data or statistical information. In practice, however, the actual knowledge could be inaccurate, and the data collected could be noisy, incomplete or out-dated. It is therefore important to incorporate the acquisition of data knowledge in the design of online scheduling algorithm. A promising solution is to use online/stochastic learning methods to exploit the random data to assist the decisions of EV scheduling in an iterative manner  \cite{zhang12design}\cite{majidpour15fast}. In this case, however, the learning algorithm efficiency is of paramount importance, as the EV data size could be enormous and the charging scheduling is a delay-sensitive application.

\subsection{Integration of Renewable Sources}
The integration of renewable sources brings both challenges and opportunities to the EV charging scheduling problem.
On one hand, EVs as energy storage can be used to reduce the intermittency of renewable sources, absorb the variability of load caused by renewable sources and even as energy carriers to transport energy from remote renewable sources to loads in urgent need of power supply.
On the other hand, renewable source could help reduce the fluctuation of base load and energy generation cost, especially for charging stations that own distributed renewable generators.
Then, the charging scheme should allocate energy from renewable sources to EVs in both cost-efficient and system-stability manners.
Besides, the integration of renewable energy introduces another layer of randomness in the system design, such that online algorithms now need to tackle the uncertainties from both the EVs and the renewable sources. Prediction and data mining play even more important role in improving the overall system performance.


\section{Conclusions}
In this article, we have provided an overview of efficient online charging scheduling algorithms to improve the power grid performance under different assumptions of future data knowledge. Besides, we have also highlighted some promising future research directions.
We believe that the adoption of advanced online EV charging scheduling algorithms in next-generation power grids will greatly improve their efficiency, reliability, security, and sustainability.

\end{document}